\documentclass[11pt,twoside]{article}
\usepackage{graphicx}
\newcommand{\BOLD}[1]{\mbox{\boldmath$ #1 $}}

\newcommand{\Cov}{\mbox{\rm Cov}}
\newcommand{\Var}{\mbox{\rm Var}}
\newcommand{\Rcode}[1]{}
\begin{document}
\title{Using interpolation to reduce computing time for analysis of
large but simple data sets with application to design of epidemiological
studies}
\author{G.K. Robinson and L.M. Ryan}
\maketitle

\section*{Abstract}
One way to investigate the precision of estimates likely to result from
planned experiments and planned epidemiological studies is to simulate a
large number of possible outcomes and analyse the sets of possible
results.  This appears to be computationally expensive for some
multi-stage designs, so choice of designs is instead based on
theoretical derivation of expected information.
This paper shows that for some types of studies
the analysis of large numbers of simulated
outcomes can be achieved more rapidly by making use of interpolation.

\subsection*{Keywords}
Experimental design; interpolation; maximum likelihood;
multi-stage epidemiological studies

\section{Introduction}
The general problem that motivated this work is the design of
cost-effective large epidemiological studies.  More specifically,
consider investigating the health effects of environmental exposures
when the cost of accurately measuring environmental exposure is large.
Two-stage designs are often practical, with environmental exposure being
measured to greater precision in the second stage than in the first
stage.

We will consider in detail the first hypothetical two-stage trial
presented in Table~1 of Morara \textit{et al}
(2007) which is intended
for assessing the relationship between pesticide exposure and
autism.  The first stage of the study has 63350 people.  Their autism
status is assessed without error and their pesticide exposure is
measured by a method which has correlation 0.3 on a logarithmic scale
with the true pesticide exposure.  In the second stage of the study, 219
individuals are chosen at random and their true pesticide exposures are
measured accurately.

Our approach to studying the usefulness of this experimental design is
to simulate a large number of sets of data, to estimate the parameters of the model
using each set of data, and to summarize the performance of the
experimental design by looking at the distribution over simulations of
parameter estimates and their standard errors.

The logarithm of true pesticide exposure, $Z$, was taken to have a
standard normal distribution.  The logarithm of the
approximate measure of pesticide exposure, $Z_A$, was computed as
$$Z_A=Z+\varepsilon$$
where $\varepsilon$ has a normal distribution with mean zero and
variance 91/9.
This variance has been chosen so that the correlation of $Z_A$ and $Z$ is
$\Cov(Z_A,Z)/\sqrt{\Var(Z_A)\Var(Z)} = 1/\sqrt{100/9} = 0.3$.
The response variable, $Y$, is one if a person is autistic and zero
otherwise.  The logit of the probability of autism is $(Z-4.733)/0.693$.
This relationship was chosen so that a unit increase in the natural
logarithm of pesticide exposure increases the odds of autism by a factor
of 2 and the overall risk of autism is 3/1000.

After each set of data is simulated, a model of the same form as
the one used to generate the data is fitted by maximum
likelihood.
There are seven parameters: the mean
and standard deviation of the true pesticide exposure, the mean and
standard deviation of approximate pesticide exposure, the correlation
between true and approximate pesticide exposure, and the location and
scale parameters of the logistic relationship between true pesticide
exposure and the probability of autism.
These parameters are collectively denoted by $\BOLD{\theta}$.

For the 219 people participating in the second stage of the study,
their contribution to the log likelihood of $\BOLD{\theta}$
is the logarithm of the density of $Y$ given $X$ which can
be easily computed using the probability density of a logistic distribution.
For the $63350-219=63131$ people not participating in the second
stage of the study, their contribution to the log likelihood of $\BOLD{\theta}$
is the logarithm of the density of
$Y$ given $Z_A$.  This density can be computed by integrating the
density of $Y$ given $Z$ over the distribution of $Z$ given $Z_A$.  This
integration can be done to sufficient accuracy by
using 8-point Hermite integration.

\Rcode{ 219+63131* 8	}
If we compute each of the logistic densities separately, then there are
$219+63131\times 8 = 505267$ of them to be computed for each evaluation of the log likelihood.
The log likelihood will be evaluated many times during the process of estimating
$\BOLD{\theta}$ by maximum likelihood.
We expect to need to do this for many simulated data sets for each of many possible
experimental designs.
The total computer time required to do this would be inconveniently large.

\section{The basic idea}
The characteristic of this situation that we can exploit to speed up the
computations
is that, for given $\BOLD{\theta}$, the
contributions to the log
likelihood for the people not participating in the second stage of the study
are a function of only a single number, $Z_A$.
For large data sets, the total of these contributions can generally be
computed much more efficiently by using piecewise polynomial interpolation than by computing each of them directly.

For Lagrange interpolation,
Abramowitz and Stegun (1965, formula 25.2.2) tells us that
given nodes $x_0$, $x_1$, \ldots $x_n$ and corresponding function
values $f_0$, $f_1$, \ldots $f_n$, the interpolated value at $x$ is
$\sum_{i=0}^n l_i(x)f_i$ where
$$l_i(x)=\frac{(x-x_0)\ldots (x-x_{i-1})(x-x_{i+1})\ldots (x-x_n)}{
	(x_i-x_0)\ldots (x_i-x_{i-1})(x_i-x_{i+1})\ldots (x_i-x_n)}.$$
For instance, if the nodes are taken to be $x_i= i$ for $i=0, 1, \ldots,
19$ then for $x=4.5$ the weights, $l_i(4.5)$, are
$0.00001019143$, $-0.0002489622$, $0.003136923$, $-0.0296265$,
$0.355518$, $1.066554$, $-0.8295419$, $0.9243467$, $-0.9903715$,
$0.9414643$, $-0.770289$, $0.533277$, $-0.3081156$, $0.1463898$,
$-0.05613442$, $0.01692943$, $-0.003864326$, $0.0006273847$,
$-0.00006454575$ and $0.000003162859$, respectively.
\Rcode{
SetLagrangeX(0:19)
SetLagrangeInterpWts(4.5)
LgInterp$WTS
}
This method of interpolation is equivalent to putting a polynomial of
order $n$ through the interpolation nodes and evaluating this
polynomial at $x$.

We have generally used equally-spaced interpolation nodes.
Two well-known reasons for being wary of polynomial interpolation do not
stop this procedure from being useful.
\begin{enumerate}
\item When high-order polynomials are fitted
to data, the coefficients of the fitted polynomials are very
sensitive to the function values.  This is not a problem because we
only make use of the fitted values; and these are usually not particularly sensitive to the
function values, as can be seen by looking at Lagrange weights like the
ones above.  For
instance, if the value $f_{19}$ was increased by 1 then the interpolated
value at $x=4.5$ would be increased by only $0.000003162859$, though
the coefficients of the fitted polynomial would be dramatically changed.

\item
Increasing the number of equally-spaced nodes does not guarantee that
interpolation becomes more accurate, as is well known for Runge's
function $f(x)=1/(1+25x^2)$ over the
range from $-1$ to $+1$.
This potential problem is overcome by using piecewise polynomial
interpolation (rather than using a single interpolation formula over the entire range
of interest), and by allowing the interpolation nodes to have a range greater than the
range over which interpolation is required.
For instance, to interpolate Runge's function
we might use the twenty nodes $-1.06$,  $-1.04$,
$-1.02$,  $-1.00$, \ldots, $-0.68$ for computing interpolated values for
$x$ in the range from $-1.00$ to $-0.74$; and use seven other regions
of interpolation transposed to the right by multiples of 0.26 from this
range.
This set of eight polynomials fitted to overlapping sets of 20 nodes
gives a maximum interpolation error less than $10^{-8}$.  The
maximum interpolation could be further reduced by reducing the spacing of
interpolation nodes while keeping the order of interpolation constant.
\end{enumerate}

Suppose that we wish to optimize a function of the form
$S(\BOLD{\theta})=\sum_{i=1}^N g(x_i|\BOLD{\theta}) $ over some
possibly multi-dimensional parameter $\BOLD{\theta}$.
In the applications of interest, $g$ is usually log likelihood or
something similar.
Rather than computing $g(x_i|\BOLD{\theta})$ separately for each
$x_i$, first compute $g(R_j|\BOLD{\theta})$ for a set
of $M$ interpolation nodes, denoted $R_j$.
Provided that the interpolation nodes are appropriately chosen,
the values $g(x_i|\BOLD{\theta}) $ can all be closely approximated by interpolation, so
we can compute weights, $w_{ij}$, such that
$g(x_i|\BOLD{\theta}) \approx \sum_{j=1}^M w_{ij} g(R_j|\BOLD{\theta}).$
Now
\begin{equation}S(\BOLD{\theta}) \approx \sum_{i=1}^N\sum_{j=1}^M w_{ij} g(R_j|\BOLD{\theta})
= \sum_{j=1}^M\sum_{i=1}^N w_{ij} g(R_j|\BOLD{\theta})
    =\sum_{j=1}^MW_j g(R_j|\BOLD{\theta})\end{equation}
where the total weights associated with the interpolation nodes are 
$W_j=\sum_{i=1}^N w_{ij}$.
The $W_j$ do not depend on $\BOLD{\theta}$, so computing
the approximation $\sum_{j=1}^MW_j g(R_j|\BOLD{\theta})$ requires only $M$
computations of the function $g$, whereas the direct computation of
$S(\BOLD{\theta})$ requires $N$ computations of the function $g$.
This is a substantial saving, since $M<\!<N$.
Furthermore, when analysing simulated data as part of the design of a
proposed large experimental study the number of interpolation nodes, $M$,
does not need to be made larger as the size of the study, $N$, is increased.

\section{Computing practicalities}
These computations can make use of a module which is a function that takes as inputs
the spacing and range of a set of
interpolation nodes and a set of nodes, $\{x_j\}$, at
which the unspecified function $g$ is to be evaluated, and returns the set of
interpolation nodes, $\{R_j\}$, and the sums of
interpolation weights associated with these interpolation nodes,
$W_j = \sum_{i=1}^N w_{ij}$.
With many interpreted
systems, code for this key initial computation might be written in a compiled language such as
\texttt{C} or \texttt{Fortran} and made available as a dynamically linked module.

In practice, one of the most difficult tasks is choosing the
spacing between interpolation nodes.  A relevant theoretical result is
that for Lagrangian interpolation the interpolation error at $x$
between $x_1$ and $x_M$ based on function values
$f(x_1)$, $f(x_2)$, \ldots\ $f(x_M)$, can
be expressed in the form
$$(x-x_1)(x-x_2)\ldots(x-x_M)\frac{f^M(\xi)}{M!}$$
for some point $\xi$ such that $x_1 < \xi < x_M$.  See, for instance,
Abramowitz and Stegun (1965, formula 25.2.3) or 
Dahlquist and Bj\"{o}rk (2008, theorem 4.2.3).
The practical importance of this formula is that the absolute
interpolation error often behaves roughly like the $M$th power of the
spacing between interpolation nodes.  For interpolation making use of
$M$ interpolation nodes, if the absolute interpolation
error for node spacing $h$ is
$\varepsilon_1$ and it is desired to achieve absolute interpolation
error $\varepsilon_2$ then $h$ should be changed to be
$h(\varepsilon_2/\varepsilon_1)^{1/M}$ or smaller.

When applying this procedure, it is important to obtain several
estimates of the interpolation error being achieved for many different $x$ values,
in order to reduce the risk that the apparent absolute estimated
interpolation error is much smaller than is typical for similar $x$ values.

In order to reduce rounding errors, it is desirable that the absolute values of
the weights, $w_{ij}$, not be too large.
When using equally-spaced interpolation nodes, this can be achieved by
not using interpolation formulae for points near the extreme nodes.
A simple rule when using 20 equally-spaced
interpolation nodes is to ensure that there are three interpolation nodes smaller than all of
the $x_i$ and three interpolation nodes larger than all of
the $x_i$.
This rule ensures that the sum of absolute interpolation weights does not exceed
72.8.
\Rcode{
# Find largest sum of absolute interpolation weights
SetLagrangeX(0:19)
f <- function(x){
	SetLagrangeInterpWts(x)
	return(sum(abs(LgInterp$WTS)))
}
optimize(f, lower = 2.1, upper = 2.9, maximum = TRUE)
}

Our preference for equally-spaced interpolation nodes rather than
Chebyshev nodes was based on comparing the interpolation accuracy
achieved when the two methods were used for piecewise polynomial
interpolation with the same number of nodes per unit length.  For a set
of 20 standard Chebyshev nodes, namely $\cos(\pi/40)$, $\cos(3\pi/40)$, \ldots,
$\cos(39\pi/40)$ over the range from $-1$ to $+1$, there are 10 nodes
per unit length.  The nodes for the interpolation over one region are
completely distinct from the nodes used over the adjacent regions.

For equally-spaced interpolation nodes with 10 nodes per unit length, the nodes
might be at integer multiples of 0.1.  We
might use sets of 20 nodes such as those at $-1.0$, $-0.9$, \ldots,
$0.9$ for interpolation over ranges of length 1.3, such as the range from
$-0.7$ to $0.6$ in this case, if we use
the rule suggested above.  More
accurate interpolation was achieved using the equally-spaced
interpolation nodes.
The greater accuracy is achieved essentially because
(when the number of nodes per unit length is the same)
the regions used for
piecewise polynomial interpolation with equally-spaced interpolation
nodes are smaller than the regions used with Chebyshev nodes.  

An alternative comparison was based on using piecewise polynomial
interpolation regions for equally-spaced nodes 
of the same size as the regions used for Chebyshev interpolation.
Interpolation over the
range from $-1$ to $+1$ making use of the 27 equally spaced nodes
between $-1.3$ and $+1.3$
was even more accurate.  Essentially,
equally-spaced interpolation nodes give better interpolation accuracy
for a given number of nodes per unit length because the interpolation in
each region of the piecewise polynomial interpolation is able to make
use of the nodes from adjacent regions.

\Rcode{
# Compare 20 point interpolation with equally-spaced nodes and Chebyshev nodes
#	at comparable density
source("LagrangeInterpolation.R")
chebyshev.std <- .5*cos(pi*(2*(1:20) -1)/40)
f <- function(x) sin(5*x-.4)
low <- -1
high <- 1
h <- (high-low)/20
trial.x <- seq(from=low+.001*h, to=high-.001*h, length=1000)
chebyshev.x <- .5*(low+high)+chebyshev.std*(high-low)
trial.y <- f(trial.x)
chebyshev.y <- LagrangeInterp(chebyshev.x, f(chebyshev.x), trial.x)
mean(abs(chebyshev.y - trial.y))
plot(trial.x, chebyshev.y - trial.y, type="l")

# Comparison using regions of the same length (different order)
esp.x <- seq(from=low-3*h, to=high+3*h, by=h)
esp.y <- LagrangeInterp(esp.x, f(esp.x), trial.x)
mean(abs(esp.y - trial.y))
plot(trial.x, esp.y - trial.y, type="l")

esp2.x <- seq(from=low, by=h, length=20)
trial2.x <- seq(from=low+3.001*h, to=low+15.999*h, length=1000)
trial2.y <- f(trial2.x)
esp2.y <- LagrangeInterp(esp2.x, f(esp2.x), trial2.x)
mean(abs(esp2.y - trial2.y))
plot(trial2.x, esp2.y - trial2.y, type="l")
}

\section{Continuation of hypothetical epidemiological
study on the link between pesticide exposure and autism}

For the hypothetical two-stage trial assessing
the relationship between pesticide exposure and autism which was
discussed in the Introduction, we simulated 1000 different data sets
in order to study the average performance of a trial with 63350 people in
the first stage and 219 in the second stage.  This option was specifically listed in
Table 1 of Morara \textit{et al} (2007), with the claim based on the
expected information matrix that this design
has ``80\% power, assuming a two-sided test at significance level 0.05''.

Our search for the maximum likelihood estimates for the seven parameters
was done using a logarithmic scale for the standard deviations and a
$\tanh(\rho)$ scale for the correlation, $\rho$, between $Z$ and $Z_A$
so that the search space did not need to be constrained.
Standard errors of the estimates were computed using the square roots
of diagonal elements of the inverses of the hessian matrices found at the
maximum likelihood estimates.  

Out of the 1000 simulated data sets, there were two cases where the
hessian matrix at the maximum likelihood estimate was not positive definite.
Other evaluation of the usefulness of the experimental design concentrated
on the scale parameter of the logistic relationship between true pesticide
exposure and the probability of autism.
There were only 15 cases where this parameter was not larger than 1.96 times its
estimated standard error.
Therefore the power of the test is estimated by the simulations to be 98.3\%.

This is much larger than the 80\% power claimed by Morara \textit{et al} (2007).
The difference in power appears to be largely due to the fact that we have
estimated the standard error using the observed hessian matrix at
the maximum likelihood estimate, whereas
Morara \textit{et al} (2007) used the expected hessian matrix at the true
parameter values.

This is using \texttt{R} except for the computation of the sums of
interpolation weights being in \texttt{C}, the 1000 simulations took 453 seconds in total
on a one-year-old PC running at 3GHz when the interpolation-based methodology was used.
For comparison, a single simulation without using interpolation took 93 seconds.
The interpolation-based methodology is approximately 200 times faster.

\section{Normal mixture models}
As another illustration of the use of interpolation, consider the problem of
estimating the five parameters to describe a
two-component mixture of normal distributions.
Simulated data was a million simulated numbers drawn from
a mixture of a standard normal distribution with weight 30\% and 70\% weight for a
normal distribution with mean 0.4 and standard deviation 1.3.

The computer time to fit the five parameters (two means, two standard
deviations and a proportion) by maximum likelihood in the language \texttt{R}
using a directly computed likelihood was 258 seconds.
Using 20-node interpolation with  spacing $h=0.15$ between equally-spaced nodes,
the time to compute the sums of interpolation
weights using a function written in \texttt{C} and called from
\texttt{R} was 0.23 seconds and the time to fit the model was 0.08 seconds.
The answers obtained by the two methods agreed to about 8 significant figures,
which is
much more precision than necessary given that the standard errors in the
estimates of the parameters are of the order of 1\%.
Using $h=0.2$ and using $h=0.1$ took similar amounts of computer
time and gave virtually the same solution.

\section{Discussion}
The change in order of summation in Equation~(1) is the key to the
computational efficiency of the new methodology.  It allows the total of
a large number of terms to be computed without computing any of the
individual terms.

For design of large experimental trials, the methodology proposed is
more useful than calculation of the expected information matrix because
it allows many aspects of variation over the range of possible
experimental outcomes to be investigated.
It allows estimation of the probability that there will be insufficient
information to estimate all of the required parameters and estimation of
the distribution over the range of possible experimental outcomes
of standard errors of parameter estimates.
We believe that it is a useful tool for the testing of possible designs
for epidemiological trials.

However, this methodology is only useful when there is a single continuous
response variable and at most a small number of categorical predictor
variables.
It is unlikely to be useful for analysis of epidemiological trials
after the data are obtained, because the models fitted at that time will generally be
more complicated.  There will often be additional
continuous predictor variables such as age, weight, height and blood pressure.
Also, there will often be a large number of categorical predictor
variables.

\section{Acknowledgement}
This research was partially supported by a grant from the National
Institutes of Health of the United States of America.

\section*{References}
Abramowitz, M. and Stegun, I.A. (1965).  Handbook of Mathematical Function with Formulas, Graphs, and Mathematical Tables. Dover.

%Fr\"{o}berg, C-E. (1969).  Introduction to Numerical Analysis. Second Edition.
%Addison-Wesley.

Dahlquist, G and Bj\"{o}rk, \AA. (2008).
Numerical Methods in Scientific Computing.
Society for Industrial and Applied Mathematics.
% Hargrave-Andrew  518.2 D131N 2008

Morara, M., Ryan, L, Houseman, A. and Strauss, W. (2007). Optimal design
for epidemiological studies subject to designed missingness.
Lifetime Data Analysis 13, 583--605.

\end{document}